\documentclass[12pt]{article} 
\newcommand{\mathbb}{\cal}
\newcommand{\M}{{\cal M}}
\newcommand{\text}{\rm }
\begin{document}

\begin{center}
{\Large \bf The Fermion Doubling Problem and 
Noncommutative Geometry II}\\
\bigskip

{\bf A. P. Balachandran\footnote{bal@suhep.phy.syr.edu}},\\
{\it Physics Department, Syracuse University, \\
Syracuse,N.Y.,13244-1130, U.S.A.}\\
{\bf T. R. Govindarajan\footnote{trg@imsc.ernet.in}},\\
{\it Institute of Mathematical Sciences,\\
Chennai 600 113, India}\\ 
{\bf B. Ydri\footnote{ydri@suhep.phy.syr.edu}}\\ 
{\it Physics Department, Syracuse University, \\
Syracuse,N.Y.,13244-1130, U.S.A.}\\
\end{center}
\vskip.5cm
\begin{abstract}
In our previous paper (hep-th/9911087), we proposed 
a resolution for the fermion
doubling problem in discrete field theories based on the fuzzy
sphere and its cartesian products. In this paper after a review 
of that work, we bring out its relationship to the 
Ginsparg-Wilson approach.
\end{abstract}

\bigskip
The nonperturbative formulation of chiral gauge theories 
is  a long standing programme in particle physics. 
It seems clear that one should regularise
these theories with all symmetries intact.
There is a major problem associated with conventional lattice
approaches  to this programme, with roots in topological features:
The Nielsen-Ninomiya
theorem \cite{NN} states that if we want to maintain chiral symmetry, then under
plausible assumptions, one cannot avoid the doubling of fermions in the 
usual lattice formulations.

Recently a novel approach to discrete physics has been developed. 
It works with quantum fields
on a ``fuzzy space'' ${\M}_{F}$ obtained by treating the underlying 
manifold ${\M}$ as a phase space and quantising it
\cite{madore,gropre,grklpr1,grklpr2,grklpr3,watamura1,watamura2,frgrre}.
Topological features, chiral anomalies and $\sigma-$ models have been 
successfully developed  in this framework \cite{grklpr1,bbiv,bal},using the cyclic cohomology of Connes\cite{connes,coquereaux}.
 
In this paper we first review the solution 
of the fermion doubling problem discussed in hep-th/9911087 
and then  clarify its relationship to the conventional
Ginsparg-Wilson approach.
An alternative approach can be found in \cite{grklpr1}. 
There have also been important 
developments \cite{gw,luscher} in the theory of chiral fermions and 
anomalies in the usual lattice formulations. 
We will show that there are striking relationships between our approach
and these developments.

Quantisable adjoint orbits of compact semi-simple Lie groups seem 
amenable to the full fuzzy treatment and lead to manageable finite 
dimensional matrix models for quantum fields . There are two such 
manifolds in dimension four, namely $S^2{\otimes}S^2$ and 
$ {\mathbb C}{\mathbb P}^2 $. 
Our methods readily extend to $S^2{\otimes}S^2$. 
They are not anticipated to encounter obstructions for 
$ {\mathbb C}{\mathbb P}^2 $ as well. But we have not yet fully worked out 
its noncommutative geometry. The published work of Grosse and 
Strohmaier \cite{grostr} on $ {\mathbb C}{\mathbb P}^2 $ gives their 
description of fuzzy $4d$ fermions.

A sphere $S^2$ is a submanifold of ${\mathbb R}^3$:
\begin{equation} 
S^2=\langle \vec{n} \in {\mathbb R}^3: \sum_{i=1}^3 n_i^2=1 \rangle. 
\end{equation}
If $\hat{n}_i$ are the coordinate functions on $S^2$,
$\hat{n}_i(\vec{n}) = n_i$, then $\hat{n}_i$ commute and the algebra
${\cal A}$ of smooth functions they generate is commutative.
In contrast, the operators $x_i$ describing $S_F^2$ are
noncommutative:
\begin{equation} 
[x_i, x_j] = \frac{i \epsilon_{ijk} x_k}{[l(l+1)]^{1/2}}, \quad
\sum_{i=1}^3 x_i^2 = {\bf 1}, \quad l \in \{\frac{1}{2}, 1,
\frac{3}{2} \ldots \}.
\end{equation}
The $x_i$ commute and become $\hat{n}_i$ in the limit 
$l \rightarrow \infty$. If $L_i =
[l(l+1)]^{1/2}x_i$, then $[L_i, L_j] = i \epsilon_{ijk}L_k$ and 
$\sum L_i^2 = l(l+1)$, so that $L_i$ give 
the irreducible representation (IRR)
of the $SU(2)$ Lie algebra for angular momentum $l$. $L_i$ or $x_i$
generate the algebra $A=M_{2l+1}$ of $(2l+1) \times (2l+1)$ matrices.

Scalar wave functions on $S^2$ come from elements of ${\cal A}$. In a
similar way, elements of $A$ assume the role of scalar wave functions
on $S_F^2$. A scalar product on $A$ is $\langle \xi, \eta \rangle = Tr
{\xi}^{\dagger} \eta$. $A$ acts on this Hilbert space by left- and
right- multiplications giving rise to the left and right- regular
representations $A^{L,R}$ of $A$. For each $a \in A$, we thus have
operators $a^{L, R} \in A^{L,R}$ acting on $\xi \in A$ according to
$a^L \xi = a \xi, a^R \xi = \xi a$. [Note that $a^L b^L = (ab)^L $ while $a^R
b^R = (ba)^R$.] We assume by convention that elements of $A^L$ are to
be identified with fuzzy versions of functions on $S^2$.
There are two kinds of angular momentum operators 
$L_{i}^{L}$ and $-L_{i}^{R}$. The orbital angular momentum operator, 
which should annihilate ${\bf 1}$,
is ${\cal L}_i = L_i^L - L_i^R$. $\vec{\cal L}$ plays the role of the
continuum $-i(\vec{r} \times \vec{\nabla})$.

The following two Dirac operators on $S^2$ have occurred 
in the fuzzy literature:
\begin{equation} 
{\cal D}_1 = \vec{\sigma}. [-i(\vec{r} \times \vec{\nabla})] + {\bf
1},\qquad 
{\cal D}_2 = \epsilon_{ijk}\sigma_i \hat{n}_j {\cal J}_k,
\end{equation}
where 
\begin{eqnarray} 
{\cal J}_k& =& [-i(\vec{x} \times \vec{\nabla})]_k + \sigma_k/2\nonumber \\
& =& \rm{Total~ angular~ momentum~ operators~} .
\end{eqnarray}
There is a common chirality operator $\Gamma$ anticommuting with both:
\begin{equation}	             
\Gamma = \vec{\sigma}.\hat{n} = \Gamma^{\dagger}, \quad \Gamma^2 =
{\bf 1},\qquad 
\Gamma {\cal D}_\alpha + {\cal D}_\alpha \Gamma =0.\label{g2}
\end{equation} 
These Dirac operators {\it in the continuum} are unitarily
equivalent,
\begin{equation}
{\cal D}_2 = \exp{(i \pi \Gamma/4)} {\cal D}_1 \exp{(-i \pi \Gamma/4)}
~~=i{\Gamma}{\cal D}_1,\label{gamma}
\end{equation}
and have the spectrum $ \{ \pm (j+1/2): j \in \{1/2,
3/2,\ldots  \} \}$, 
where $j$ is total angular momentum 
(spectrum of $\vec{\cal J}^2 =\{j(j+1)\}$). 

Since $|{\cal D}_{\alpha}|$ $(\equiv $ positive square root of 
${\cal D}_{\alpha}^2$ $)$ for both ${\alpha}$ share the same spectrum and 
rotational invariance, 
$ |{\cal D}_{1}|=|{\cal D}_{2}| $. 
Further being multiples of unity for each fixed $j$, they commute 
with the rotationally invariant ${\Gamma}$. 
As they are invertible too, we have the important identity 

\begin{equation}
\Gamma~=~i{\frac{{\cal D}_1}{|{\cal D}_1|}}{\frac{{\cal D}_2}
{|{\cal D}_2|}}\label{gama}.
\end{equation}
The discrete version of ${\cal D}_1$ is: 
\begin{equation} 
D_1 = \vec{\sigma}. \vec{\cal L} + {\bf 1} ,
\end{equation} 
while
\begin{eqnarray}
\text{Spectrum of}\,\,D_1 &=&  \left\{ \pm (j+\frac{1}{2}): j \in
                              \{ \frac{1}{2}, \frac{3}{2}, \ldots
                              2l-\frac{1}{2} \} \right\} \nonumber \\  
                        &\cup& \left\{ (j+\frac{1}{2}):
                              j=2l+\frac{1}{2} \right\}.\label{specd1}
\end{eqnarray}
It is easy to derive (\ref{specd1}) by writing 

\begin{eqnarray}
D_1&=& \vec{J}^2 -
\vec{\cal L}^2 - \left(\frac{\vec{\sigma}}{2}\right)^2 + {\bf 1},~~ 
\left(\frac{\vec{\sigma}}{2}\right)^2 = \frac{3}{4} {\bf 1},\\
{\cal L}_k +\frac{{\sigma}_k}{2}&=&J_k \label{angmom}\\
&=& ~\text{Total} ~\text{angular} ~\text{momentum} ~\text{operators}. 
\end{eqnarray}
We let $j(j+1)$ denote the eigenvalues of ${\vec{J}}^2$.
Then for $\vec{\cal L}^2 = k(k+1), k \in \{0, 1, \ldots 2l \}$, if
$j=k+1/2$ we get $+(j+1/2)$ as eigenvalue of $D_1$, while if $j=k-1/2$ we get
$-(j+1/2)$. The absence of $-(2l+1/2)$ in (\ref{specd1}) is just
because $k$ cuts off at $2l$.
(The same derivation works also for ${\cal D}_1$).

The discrete version of ${\cal D}_2$ is :
$D_2 = -\epsilon_{ijk}\sigma_i x^L_j J_k = \epsilon_{ijk}\sigma_i 
x_j^L L_k^R$.
$D_2$ is  no longer unitarily equivalent to $D_1$, its spectrum  
is given in \cite{watamura1,watamura2}.

The first operator has been used extensively by Grosse et al
\cite{grklpr1,grklpr2,grklpr3} while the second was 
first introduced by Watamuras
\cite{watamura1,watamura2}.
It is remarkable that the eigenvalues (\ref{specd1}) coincide {\it exactly} with
those of ${\cal D}_\alpha$ upto $j=(2l-1/2)$. 
In contrast $D_2$  
has zero modes when $j~=~2l~+\frac{1}{2}$ and very small eigenvalues 
for large values of $j$, both being absent for ${\cal D}_{\alpha}$ .  
So $D_1$ is a better approximation to ${\cal D}_\alpha$.

But $D_1$ as it stands admits no chirality operator
anti-commuting with it. This is easy to
see as its top eigenvalue does not have its negative
in the spectrum. Instead  $D_2$ has the nice 
feature of admitting a chirality operator: the eigenvalue
for top $j$, even though it has no pair, is  exactly zero. 
So the best fuzzy Dirac operator has to combine the good
properties of $D_1$ and $D_2$. 
We suggest it to be $D_1$ after projecting out its top
$j$ mode. We will show that it then admits a chirality
with the correct continuum limit .

The chirality operator anticommuting with $D_2$ and squaring to ${\bf 1}$
in the {\it entire} Hilbert space is
\begin{equation}
\gamma_R = \gamma_R^{\dagger} = -\frac{\vec{\sigma}.{\vec{L}}^R
-1/2}{l+1/2},~
\gamma_R^2 = {\bf 1}\label{gR}.
\end{equation}
An interpretation of ${\gamma}_R$ is that $(1{\pm}{\gamma}_R)/2$ 
are projectors to subspaces where $(-\vec{L}^R+\vec{\sigma}/2)^2$ 
have values $(l{\pm}\frac{1}{2})(l{\pm}\frac{1}{2}+1)$ \cite{bbiv}.
The following identity is easily shown :
\begin{equation}
[D_1,\gamma_R]~=-~2~i~{\lambda}~D_2;~~ 
{\lambda}~=~\sqrt{1-\frac{1}{(2l~+~1)^2}}~.
\label{identity}
\end{equation}
Now $D_{\alpha}^2$ and $ |D_{\alpha}|(\equiv $ nonnegative 
square root of $D_{\alpha}^2)$ are multiples of identity for 
each fixed {\it j}, and ${\gamma}_R$ commutes with ${\vec{J}}$. 
Hence they mutually commute :

\begin{equation}
[A,B]~=~0~ \text{for}~ A~,B~=~D_{\alpha}^2~ , 
|D_{\alpha}|~ \text{or}~ {\gamma}_R~ .
\end{equation}
Therefore from (\ref{identity}),
\begin{equation}
\{D_1,D_2\}=\frac{1}{2i{\lambda}}[D_1^2,{\gamma}_R]=0 .
\end{equation}
In addition we can see that 
$[D_{\alpha}^2,D_{\beta}]=[|D_{\alpha}|,D_{\beta}]=0$.
If we define
\begin{eqnarray}
\epsilon_\alpha &=& \frac{D_\alpha}{|D_\alpha|}
 \quad \text{on~the~subspace}~ V~ 
\text{with}\;j \leq 2l-1/2,
 \nonumber \\ 
 &=& 0 \quad \text{on~the~subspace}~ W~ 
\text{with}\;j=2l+1/2,
\end{eqnarray}
it follows that 
\begin{equation}
e_1={\epsilon}_1~ , e_2={\epsilon}_2~ , e_3=i{\epsilon}_1{\epsilon}_2 
\label{e1e2e3}
\end{equation}
generate a Clifford algebra on $V$. 
That is, if $P$ is the orthogonal projector on $V$,
\begin{eqnarray} 
P \xi &=& \xi, \quad \xi \in V, \nonumber \\
      &=& 0,  \quad \xi \in W,
\label{pdef}
\end{eqnarray}
then $\{e_{\alpha},e_{\beta}\}=2{\delta}_{{\alpha}{\beta}}P$.

All this allows us to infer that $\{e_3,D_1\}=0$ so that 
$e_3$ is a chirality operator for either $D_1$ or its 
restriction $PD_1P$ to $V$. In addition, {\it in view of (\ref{gama}), 
it  has the correct continuum limit as well} so that it is 
a good choice for chirality in that respect too.

A unitary transformation of $e_3$ and $D_1$ will not disturb 
their nice features. Such a transformation bringing $e_3$ to 
${\gamma}_R$ on $V$ is convenient. It can be constructed as 
follows. $e_{\alpha}$ and ${\gamma}_R$ being rotational scalars 
leave the two-dimensional subspaces in $V$ with fixed values  of 
$\vec{J}^2$ and $J_3$ invariant. On this subspace, $e_{\alpha}$ 
and unity form a basis for linear operators, so ${\gamma}_R$  is 
their linear combination. As $e_1$,$e_3$ and ${\gamma}_R$ 
anticommute with $e_2$,and all square to ${\bf 1}$, in this subspace, 
we infer that ${\gamma}_R$ is a transform by a unitary operator 
$U=exp(i{\theta}e_2/2)$ of $e_3$ in each such subspace. 
And ${\theta}$ can depend only on $\vec{J}^2$ by rotational invariance. 
Thus we can replace $PD_1P$ and $e_3$ by the new Dirac and 
chirality operators
\begin{eqnarray} 
D &=& e^{(i \theta (J^2) \epsilon_2)/2} (P D_1 P) e^{(-i \theta (J^2)
\epsilon_2)/2},\nonumber\\
{\gamma}&=&P\gamma_2 P = e^{(i \theta (J^2) \epsilon_2)/2}
(i\epsilon_1\epsilon_2) e^{(-i \theta (J^2) \epsilon_2)/2} \nonumber \\
 &=& \cos \theta (J^2) (i\epsilon_1\epsilon_2) + \sin \theta (J^2)
\epsilon_1. 
\end{eqnarray}
The coefficients can be determined by taking traces with 
${\epsilon}_1$ and $i{\epsilon}_1{\epsilon}_2$.

We have established that chiral fermions can be defined on $S^2_F$ with no 
fermion doubling, at least in the absence of fuzzy monopoles.
We later indicate how to include them as well.
\vskip2mm

\noindent{\it {These fermions are Ginsparg-Wilson}}
\footnote{This section is the joint work of 
A. P. Balachandran and Giorgio Immirzi, reproduced here with the 
gracious permission of Georgio.}:
The Ginsparg-Wilson chiral fermion has a Dirac operator $D'$
and a hermitian chirality operator $\Gamma$ squaring to unity.
$D'$ and $\Gamma$ fulfill the relations
\begin{equation}
D'^{\dagger}~=~\Gamma~D'~\Gamma,~\{~\Gamma,~D'~\}~=~aD'\Gamma D'
\end{equation}
`$a$' being lattice spacing in suitable units. Now if $\Gamma'~=~
\Gamma(a~D')~-~\Gamma$, then 
\begin{equation}
\Gamma'^{\dagger}~=~\Gamma',~
\Gamma'^{2}~=~1~ {\text{and}}~aD'~=~\Gamma(\Gamma~+~\Gamma').
\end{equation}
Conversely
given two idempotents $\Gamma$ and $\Gamma'$, we have a 
Ginsparg-Wilson pair $D'~=~\frac{1}{a}~\Gamma(\Gamma~+~\Gamma')$
and $\Gamma$. 

Our fermion on $S_F^2$ admits such a formulation, except that we 
choose $\frac{1}{a}(\Gamma~+~\Gamma')$ as the Dirac operator. Thus just like 
(\ref{gR}), we can also construct a left- 
chirality operator anticommuting with $D_2$:
\begin{equation}
\gamma_L = \gamma_L^{\dagger} = \frac{\vec{\sigma}.{\vec{L}}^L
+1/2}{l+1/2},~~
\gamma_L^2 = {\bf 1}.
\end{equation}
Then $aD_1~=~\gamma_L~+~\gamma_R$, $a~=~\frac{1}{l+1/2}$. Identification
of $\gamma_{L,R}$ with $\Gamma, \Gamma'$ establishes our
assertion.

There is a beautiful structure underlying the algebra
$K$ of the idempotents $\Gamma, \Gamma'$. It lets us  
infer certain salient
features of $D_1$, but the results transcend $S_F^2$.
We shall now briefly describe it. 

Introduce the hermitian operators 
\begin{equation}
\Gamma_{1,2}~=~\frac{1}{2}(\Gamma \pm
\Gamma'),\Gamma_3~=~\frac{i}{2}[\Gamma,\Gamma'],\Gamma_0~=~\frac{1}{2}
\{\Gamma,\Gamma'\}.
\end{equation}
Then (a) $\Gamma_m~(m\neq 0)$ mutually 
anticommute.(b) $\Gamma_0$ and $\Gamma_m^2$ commute with all $\Gamma_\lambda$
and are in the center of $K$.(c) $\Gamma_1^2~+~\Gamma_2^2 = {\bf 1}
=\Gamma_3^2~
+~\Gamma_0^2$ so that $\Gamma_\lambda$ do not have eigenvalues 
exceeding $1$  in modulus. (d) $\Gamma_1^2~=~\frac{1}{2}({\bf 1}+\Gamma_0),
\Gamma_2^2~=~\frac{1}{2}({\bf 1}-\Gamma_0)$.

The vector space we work with can be split into the direct sum
$\oplus V_j$ of eigenspaces $V_j$ of $\Gamma_0$ with distinct 
eigenvalues $\cos 2\theta_j$.  Then $KV_j~=~ V_j$. On $V_j$,
the possible eigenvalues of $\Gamma_3$ are 
${\pm}\sin 2{\theta}_j$ by (c) and
those of ${\Gamma}_{1,2}$ are ${\pm}cos{\theta}_j,{\pm}sin{\theta}_j$
by (d). Both signs do occur on $V_j$ unless $|{\Gamma}_0|$ is
${\bf 1}$ and hence ${\Gamma}_3=0$. For if ${\Gamma}_3{\neq}0$
on $V_j$,  ${\Gamma}_{1,2}$ have no zero eigenvalue and
${\Gamma}_{m}/|{\Gamma}_{m}| $ generate a Clifford algebra there.
The result follow from the identity
\begin{equation}
\frac{{\Gamma}_{n}}{|{\Gamma}_{n}|}
{\Gamma}_{m} \frac{{\Gamma}_{n}}{|{\Gamma}_{n}|}=-{\Gamma}_{m},
m{\neq}n~ on~ V_j~ for~ m,n=1,2,3.
\end{equation}
But if $|{\Gamma}_0|={\bf 1}$
and ${\Gamma}_{3}=0$ for $j=j_0$, then ${\Gamma}_1$ or ${\Gamma}_2$
is also zero on $V_{j_{0}}$ and we cannot infer that the
nonzero ${\Gamma}_{m}$ has eigenvalues of both signs there.
We can only say that its modulus $|{\Gamma}_{m}|$ has its
maximum value ${\bf 1}$ there.

In our case, $D_1$ and $D_2$ upto factors are
${\Gamma}_1$ and ${\Gamma}_3$. Previously we saw that $D_1$
admits chirality on subspaces with fixed total angular
momentum except on the subspace $W$ with maximum
total angular momentum $2l+1/2$. On $W$, $D_2$
vanishes and $|D_1|$ has its maximum value.
Also $D_1$ always admits two chirality operators
${\gamma}_{L,R}$. These results
follow from above if $j$ is identified with total
angular momentum and $V_{j_0}$ with $W$. These
identifications are consistent with the fact that
\begin{equation}
{\Gamma}_{0}=\frac{1}{2}\{{\gamma}_{L},
{\gamma}_{R}\}=\frac{a^2}{2}[\vec{J}^2 - 2l(l+1) -\frac{1}{4}].
\end{equation}

The spectrum and eigenstates of the Ginsparg-Wilson $D^{'}_1$ can 
also be found.
Thus $aD^{'}_{1}=2({\Gamma}_{1}^2 + {\Gamma}_{1}{\Gamma}_{2})$.
It can be diagonalised since
$[{\Gamma}_1^2 , {\Gamma}_1{\Gamma}_2]=0$.
On $V_j$, ${\Gamma}_1^2=cos^2{\theta}_j {\bf 1}$
and $({\Gamma}_1{\Gamma}_2)^2=-\cos^2{\theta}_{j} \sin^2{\theta}_j{\bf 1}$
or spectrum of
$aD^{'}_1$ on $V_j$ $={{1} + exp({\pm}2i{\theta}_j)} $.
It lies on a circle with center $1$ and radius $1$.
As for its eigenvectors,, we can proceed as follows.
On $V_{j_0}$, ${\Gamma}_1$ or ${\Gamma}_2$ $=0$
so that $aD'_1$ is in any case diagonal.
On $V_j$ $(j{\neq}j_0)$ , $|{\Gamma}_2|{\neq}0$ and
\begin{equation}
aD'_1=exp(-i\frac{{\Gamma}_2}{|{\Gamma}_2|}{\pi}/4)2[{\Gamma}_1^2 -
i|\sin{\theta}_j|{\Gamma}_1]exp(i\frac{{\Gamma}_2}{|{\Gamma}_2|}{\pi}/4).
\end{equation}
So if ${\Gamma}_1{\psi}_{j}^{\pm}={\pm}(\cos{\theta}_j){\psi}_{j}^{\pm}$,
then
\begin{equation}
aD'_1 exp(-i\frac{{\Gamma}_2}{|{\Gamma}_2|}{\pi}/4){\psi}_{j}^{\pm}=
[1 + exp ({\mp}\frac{\sin{\theta}_j}{|{\sin \theta}_j|}
2i{\theta}_j)](\frac{{\Gamma}_2}{|{\Gamma}_2|}){\psi}^{\pm}_j.
\end{equation}
\vskip2mm
\noindent{\it{Monopoles and Instantons}:}
In the continuum, monopoles and instantons are particular 
connection fields $\omega$
on certain twisted bundles over the base manifold ${\cal M}$. On
$S^2$, they are monopole bundles, on $S^4$ or ${\mathbb C}{\mathbb
P}^2$, they can be $SU(2)$ bundles. 

In algebraic $K$-theory, it is
well-known that these bundles are associated with projectors ${\cal
P}$ \cite{connes,coquereaux,mssv}.  ${\cal P}$ is a matrix 
of some dimension $M$ with ${\cal P}_{ij}
\in {\cal A} \equiv {\cal C}^{\infty}({\cal M})$, ${\cal P}^2 = {\cal
P} = {\cal P}^{\dagger}$. The physical meaning of ${\cal P}$ is the
following. Let ${\cal A}^M = {\cal A} \otimes {\mathbb C}^M = \{ \xi
=(\xi_1, \xi_2 \ldots \xi_M):\xi_i \in {\cal A} \}$. Then ${\cal PA}^M
= \{ {\cal P}\xi: \xi \in {\cal A}^M\}$ consists of smooth sections
(or wave functions) ${\cal P} \xi$ of a vector bundle over ${\cal
M}$. For suitable choices of ${\cal P}$, we get monopole or instanton
vector bundles. These projectors are known \cite{mssv} and
those for monopoles have been reproduced in \cite{bbiv}.

The projectors $p^{(\pm N)}$ for fuzzy monopoles of charge $\pm N$
have also been explicitly found in \cite{bbiv}. 
They act on $A^{2^N} = \{ \xi$
with components $\xi_{b_1 \ldots b_N} \in A, b_i \in \{1,2\} \}$.
Let $\vec{\tau}^{(i)}$ ($i=1, 2, \ldots N$) be commuting Pauli
matrices. $\vec{\tau}^{(i)}$ has the normal action on the index $b_i$
and does not affect $b_j$ ($j \neq i$). Then $\vec{K} = \vec{L^L} +
\sum_i \vec{\tau}^{(i)}/2$ generates the $SU(2)$ Lie algebra and
$p^{(N)}$ ($p^{(-N)}$) is the projector to the maximum (minimum)
angular momentum $k_{\text{max}}=l+N/2$
($k_{\text{min}}=l-N/2$). [$\vec{K}^2 p^{(N)} =
k_{\text{max}}(k_{\text{max}}+1) p^{(N)}$, $\vec{K}^2 p^{(-N)} =
k_{\text{min}}(k_{\text{min}}+1) p^{(-N)}$.] Fuzzy analogues of
monopole wave functions are $p^{(\pm N)} A^{2^N}$. 
When spin is included, we must enhance $p^{(\pm N)} A^{2^N}$ to
$p^{(\pm N)} A^{2^N} \otimes {\mathbb C}^2 = p^{(\pm N)} A^{2^{N+1}} =
\{\xi$ with components $\xi_{b_1 \ldots b_N, j} \in A: b_i, j \in
\{1,2 \} \}$.

The Dirac and chirality operators can be constructed for fuzzy monopoles
as well. They are described in \cite{bal}.

\noindent {\bf Acknowledgments:} 
Giorgio Immirzi, Sachin Vaidya, Xavier
Martin, Denjoe O'Connor and Peter Pre\v{s}najder offered 
us many good
suggestions during this work while Apoorva Patel told us about L\"{u}scher's 
work
\cite{luscher}. We thank them for their help. 
We are in particular thankful to Giorgio for generously 
allowing us to use his joint work with Balachandran here. The 
work of APB was supported in part by the DOE under contract number
DE-FG02-85ER40231.

\bibliographystyle{unsrt}

\end{document}